# Modeling Digital Penetration of the Industrialized Society and its Ensuing Transfiguration

Johannes VRANA [1,2][*][†] and Ripudaman SINGH [3][†]
[1] Vrana GmbH, Rimsting, Germany
[2] RIVK gGmbH, Munich, Germany
[3] Inspiring Next, Cromwell, CT, USA
* Corresponding author; Email: contact@vrana.net
† These authors contributed equally to this work

## Abstract

The Fourth Industrial Revolution, ushered by the deeper integration of digital technologies into professional and social spaces, provides an opportunity to meaningfully serve society. Humans have tremendous capability to innovatively improve social well-being when the situation is clear. Which was not the case during the first three revolutions. Thus, society has been accepting lifestyle changes willingly and several negative consequences unwillingly. Since the fourth one is still in its infancy, we can control it better. This paper presents a unified model of the industrialized ecosystem covering value creation, value consumption, enabling infrastructure, required skills, and additional governance. This design thinking viewpoint, which includes the consumer side of digital transformation, sets the stage for the next major lifestyle change, termed Digital Transfiguration. For validation and ease of comprehension, the model draws upon the well-understood automobile industry. This model unifies the digital penetration of both industrial creation and social consumption, in a manner that aligns several stakeholders on their transformation journey.

## 1 Introduction

The term "Digital Transformation" was coined in 2011 and initially defined as "the use of technology to radically improve performance or the reach of businesses" [1]. Over time the understanding of this term has evolved. Some of the popular definitions today are all given by management consulting firms with little or no regards to social acceptance. For example:

- "Digital transformation is the rewiring of an organization, with the goal of creating (business) value by continuously deploying tech at scale." [2]
- "Digital transformation can refer to anything from IT modernization (for example, cloud computing), to digital optimization, to the invention of new digital business models." [3]
- "Digital transformation is the essential bridge between the business of today and the business of tomorrow." [4]

While businesses continue to invest in digital for growth and competitive advantage, the appreciation of social engagement continues to lag. A scientific approach to understanding the change drivers and its impact appear to be missing.

Vial [5] reviewed 282 papers in 2019 to understand specific aspects of digital transformation and felt that there was still a lack of a comprehensive portrait of its nature and implications. He found 28 sources offering 23 unique definitions. He used systematic decomposition to create a new definition of digital transformation: "A process that aims to improve an entity by triggering significant changes to its properties through combinations of information, computing, communication, and connectivity technologies." Even this definition covers just the organization side of transformation.

The in-depth review of prior art [5-13] in the Supplementary Materials reveals the progress, challenges, and gaps of Digital Transformation. There is a clear void in the understanding of how digital transformation can be structured and viewed in a manner where product/service creation and market acceptance can happen in sync with changes in social norms. There are misconceptions and confusion making a difficult subject even more complex. The scope and impact of digital deserves a scientific approach.

This paper introduces a novel unified model of the industrialized society which helps identify and understand the fundamental steps of digital penetration: digitization, digitalization, and digital transformation. The multidisciplinary design thinking approach used to sketch the model is discussed in the following. The resulting model helps to differentiate the various concurrent digital penetration efforts and their goals and finally, it reveals the next step: **digital transfiguration**, which now encompasses the social consumption side of digitally driven changes.

## 1.1   Design Thinking Approach

The rapid and ever-changing landscape of technology, coupled with the increasing complexity of applications and the critical role of human involvement, necessitates a reliable approach to drive the development and adoption of digital transformation. It requires design thinking, an iterative process that delves into comprehending users, questioning assumptions, and redefining problems to uncover alternative strategies and solutions that may not be immediately evident at first glance. Moreover, design thinking not only fosters a solution-oriented mindset but also serves as a comprehensive set of practical methods for tackling challenges in a hands-on manner. According to [14] "Thinking like a designer can transform the way you develop products, services, processes—and even strategy" The prevailing application style predominantly follows an iterative loop comprising Empathize, Define, Ideate, Prototype, and Test. This foundational philosophy gives rise to numerous variations, essentially outlining a learning journey that commences with a deep understanding of the end beneficiary and culminates in a viable and effective solution.

Based on the authors experience in the digitalization and digital transformation of the inspection industry [15-18] and their debate on the current scenario in [13] they applied an industry-agnostic design thinking perspective in this paper.

1. **Empathize and Define:** An in-depth review of prior arts (see Supplementary Material) and discussions with a multitude of thought leaders helped define the gap – a systematic approach to understanding digital developments and acceptance.
2. **Ideate:** Several ideation sessions and hypothesis testing against the automobile industry [12] helped build and understand the logical model of digital penetration of the ecosystem. The scope kept increasing to include technology value proposition, process improvement, organization transformation, consumer psychology, social family structure, and hierarchy of human needs.
3. **Prototype and Validate** the model on a variety of industries: A series of conversations with business and social leaders as potential beneficiaries confirmed this approach and inspired the authors to construct a unified model, easy to visualize and communicate.

The model serendipitously evolved beyond business transformation to address the purpose around human needs and desires. This formed the connectivity with social consumption side, paving the path to a stage beyond business transformation – The ecosystem transfiguration. Unfortunately, the depth and breadth of the model make it hard to prove, by any means other than cognitive reasoning. Authors admit that it is a view of the emerging changes to the industrialized society, which comes with uncertainty and ambiguity.

## 1.2   Portrait of a Transformation

The best way to understand transformation is through an easy-to-understand real life story. On 5th August 1888, Bertha Benz drove from Mannheim to Pforzheim, Germany in a Benz Patent-Motorwagen Model 3, without telling her husband Carl Benz, asking for directions along the way, thus becoming the first person to drive an automobile over a significant distance of 106 km using wagon tracks [19]. She did this to give her husband the confidence that he had a future in making horseless carriages. She solved numerous problems along the way – found fuel at a chemist shop to run the car, used her garter as insulation material, asked a cobbler to install leather on failed brakes, added water to cool the engine at every stop, and perhaps more than anyone will ever know. The trip gave Benz(s) well-needed publicity and offered serious lessons in the development of an automobile, including the concept of test drives. Moreover, it showed the necessity for the development of basic infrastructure, like roads, gas stations, and repair shops, to make this new form of travel a reality.

That was the beginning of a revolution in transportation, initially termed the horseless carriage. This shows how far society has developed since then, with automobiles, (safer, quieter, efficient, etc.), infrastructure (highways, bridges, gas stations, repair centers, traffic controls, etc.), governance (laws and enforcement, driver's license, etc.), industrial manufacturing (automation and supply chain) and everything that seems so natural now. It is highly unlikely that Carl and Bertha Benz saw all this. That is transformation, where it becomes hard to connect with the state that existed when it all started. Just like a caterpillar transforms into a butterfly.

Similar developments can be seen in other sectors. The wired telephone created by Alexander Bell has transformed into global wireless communication using handheld devices, that do much more than just calls. The modern flying experience bears little resemblance to the first commercial flight, nor do the modern screen-dominated office environments.

Basically, any radical innovation requires a lot more active development, before it can be scaled for widespread acceptance. Anything that has the potential to transform the ecosystem, with changes in workplace and lifestyle requires new infrastructure, new skills, and new regulations to evolve concurrently. The infrastructure requires its own technology, innovation, and standardization, and when that lags, innovation stalls. Space travel is a good example of the need for a lot of infrastructure, competencies, and regulations, for it to transform into something affordably accessible to scientists and tourists.

Transformation takes a lot more than invention, a new product, or a new concept. In fact, changes to infrastructure in turn impact product innovation. The mutual dependence and evolution eventually create a completely new ecosystem. This is a long-drawn process, with no clear end.

## 1.3   Transformation Today

Digital Transformation involves the integration of digital technologies into various aspects of an organization to fundamentally transform the business of value creation. This change is viewed differently by different people because of limited experiences, diverse perspectives, narrow fields of view, and near-term business objectives; generally, correct but incomplete. Moreover, with instant communication and rapid travel, multiple changes are happening simultaneously, tied to every aspect of daily life and the workplace. The pace of change feels exponential, often leading to anxiety and stress. This complexity and ambiguity preclude people from clearly seeing the digital world of tomorrow. Just like the Benz family could not have seen the all the research and developments of the 'horse-less transportation' systems of today, an accurate comprehension of the driver-less mobility of tomorrow is challenging. A similar lack of clarity exists in every other sector — communication, education, healthcare, energy, space, and of course sustainable development.

To appreciate the complexity, the imagination of building the ecosystem on a new planet helps. Considering thousands of teams exploring and innovating at the same time, all sharing their discoveries and new technologies, accelerating every other effort in creating a new landscape, a buzzing lifestyle. Total digital transformation has the potential like landing on a new planet, bringing a different meaning to life and living.

The diversity of experiences and debatable views of the future are required for a stable revolution. But they are healthy only up to a point, beyond which the ecosystem begs for clarity, such that the infrastructure can be developed, required skills can be defined, and additional regulations are enforced; all to support the synergistic transformation to a meaningful future. Society has now reached that point with digital penetration of life and the workplace. Too many possibilities are happening simultaneously with not enough scientific thought given to the future state and enabling factors.

## 2   Industrialized Society – A Unified Model

Industrialization of society over the last few centuries can be easily justified as humans strive for assistance, prosperity, comfort, and security at various levels. This seems to be a primal need, as defined by Abraham Maslow in 1943, published as a social psychological model on human motivation [20]. Humans, demand and consume value with an insatiable appetite for new and more. The first two industrial revolutions helped most humans to move from "Physiological needs" to "Love and Belonging". The third revolution brought humanity to an economy of abundance where "Self-esteem" and freedom of lifestyle became the driving force [15]. Human needs and desires provide an incentive to invent and innovate - create value in form of new products and services. This relentless pursuit of a better lifestyle brought humanity from the hunter-gatherer era, through agricultural to the industrialized society. And with time, society transformed into a structured ecosystem, with governing controls in place. This (analog) industrialized society built the foundations for the ongoing digital revolutions.

In the present-day ecosystem, organizations and humans continuously create and consume value with discipline and structure. Leaders are driven by purpose, objectives, or incentives. Managers leverage available resources and infrastructure and conform to rules, regulations, and moral conduct. Products and services flow to consumers through various business transactions. Fig. 1 is a conceptual model of the ecosystem of industrialized society, serendipitously born during the first industrial revolution and matured during the second revolution. Digital penetration is now making it increasingly more effective and efficient.

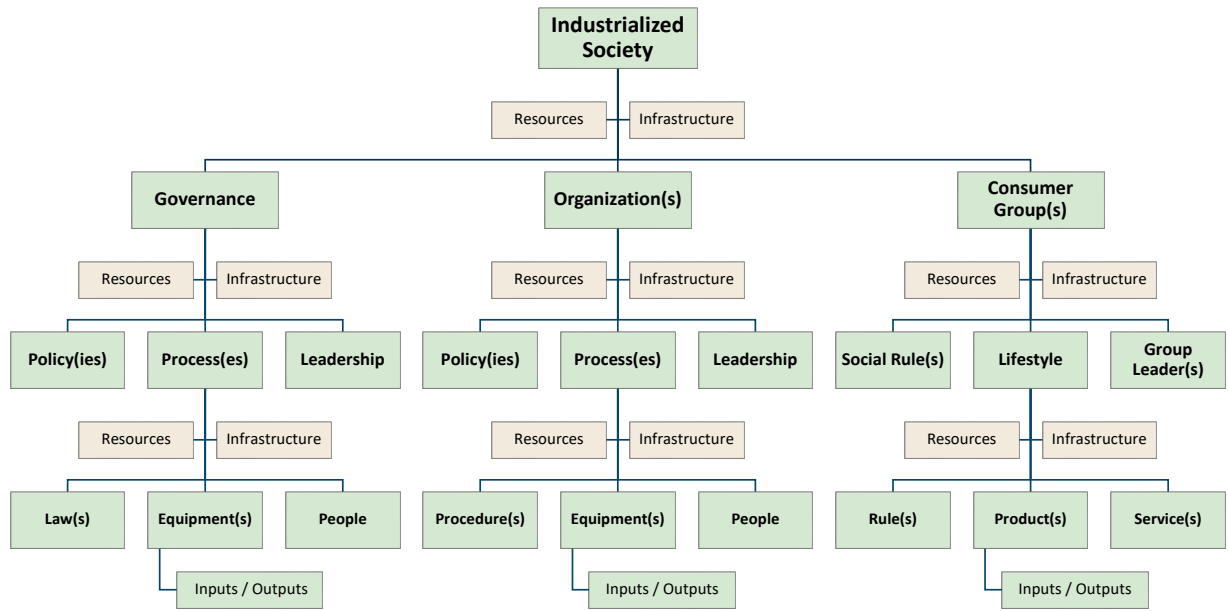

*Figure 1 **Unified model of the industrialized society:** a conceptual model of the ecosystem of industrialized society, serendipitously born during the first industrial revolution and matured during the second revolution.*

The ecosystem of the industrialized society consists of various interest groups (consumers), organizations (private, public, academic, non-profits, ...) as well as governance (governments, regulatory bodies, professional societies, ...). The flow of value requires infrastructure and enablers (discussed in detail later) which must operate within available resources and budgetary constraints. For an organization to add or create value, several processes are required, which are controlled through compliance with organizational policies, the use of company infrastructure, and are directed by the leadership, within its budgetary and resource limitations. To execute various processes, people use equipment and devices as per standard operating procedures, consume energy and other resources, and leverage facilities and logistics infrastructure. There is a similar hierarchical structure for the governance vertical as well, requiring no additional description.

Consumer groups share certain lifestyle(s), which are governed by social views/rules and are directed by group leaders. To enjoy the lifestyle, the interest groups depend on infrastructure that is either available publicly or through associations. Individuals use products and services as per family rules, consume energy and other resources, and leverage their private infrastructure.

Coming back to the automobile sector - car manufacturers use various processes to design, build, sell, and service cars. Those cars are either sold to other companies (B2B) and used as equipment or they are sold to individuals (B2C) and used as products. The government provides governance through the transportation department and state laws around driving regulations. One can map all this using the unified model presented here.

This unified model fits quite well across multiple industrial sectors since it gets down to the basics of value creation, value consumption, and the governance of value transfer at various levels.

## 3  Digital Penetration of the Industrialized Society

With the maturity of the third industrial revolution through advances in digital devices, humans are beginning to create radically new value propositions, previously not possible, within the established ecosystem represented by the unified model in Figure 1. Automation trends allow computers to run all repetitive tasks and let humans focus on creativity. Visually enhanced communications are making virtual meetings life-like. Additive manufacturing is making instant prototyping and mass customization feasible. Costs are plummeting from digital solutions. The focus has expanded from just addressing efficiency needs to including comfort and growth desires. New business models are creating possibilities that were not imaginable just two decades back. Well-established businesses like Blockbuster, Kodak, or Nokia disappeared, and so-called digitally native companies like Netflix and Uber saw unprecedented growth. Advances in computation and communication are at the core of this seismic shift.

Not all effects are positive. Along with the benefits of instant communication came the nuances of data sharing, the perils of social media, and the intensifying debate between sharing information for security and respect for individual privacy. Industrial revolutions also brought unnecessary waste and climate change. The appreciation for sustainability and sustainable development is bringing a new world order which does threaten humanity to be pushed down Maslow's hierarchy of needs. This puts consumer behavior in focus as well. The ongoing digital revolution has the potential to balance economic and ecological ecosystems. But it must be understood how to keep the rapid digital penetration under control. The unified model presented above provides a framework for such a conversation, with a view of the future.

An in-depth review of the prior art and understanding of digital transformation (included in the Supplementary Materials) identifies that the business community is heavily focused on industrial value creation with a lean mindset. The subjects needing attention for the meaningful evolution of society include social consumption, infrastructure, skills, and governance of the creation-consumption cycle.

The digital penetration of the industrialized society can be easily understood using the unified model in four transformational steps, from the bottom up: transformation of the equipment (digitization), process (digitalization), organization (digital transformation), and the industrialized ecosystem (digital transfiguration).

## 3.1 Digitization

Fig. 2 shows digitization as the change to the most fundamental element at the lowest level of the ecosystem (as depicted with the change in color from green to blue) and started with the development of (micro)electronics. This major change in equipment with digital input and output, and processing capabilities marked the onset of the digital revolution (or Industry 3.0). Examples include a digital calculator, a digital camera, a computer, a CNC machine, etc.

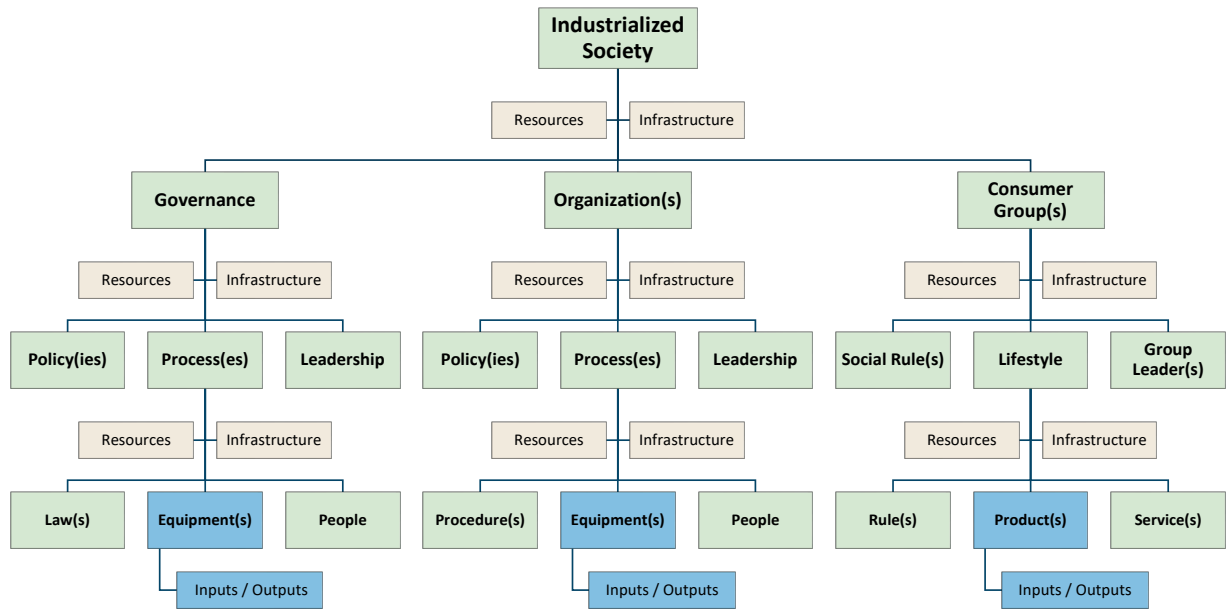

*Figure 2 **Digitization**: Transformation at equipment or device level including inputs and outputs, where relevant.*

The digital infrastructure of binary data encoding led to a universality of storage, exchange, and application at a technology level. From a consumer or business owner perspective, it offered a host of performance benefits such as miniaturization, affordability, cost, speed, and reliability of products.

For a company or a consumer, digitization can be seen as an activity or a task of replacing analog equipment with a digital device. In the context of the automotive sector simple digital manufacturing equipment (like CNC machines) came to the car manufacturers, while many things changed under the hood of the car, not obvious to the owner. For example, digitally connected input controls, speedometers, signal lamps, and other pointer indicators.

## 3.2   Digitalization

Fig. 3 shows digitalization as the change to the basic process level within organizations and governance as well as the change to the lifestyle of consumers. Isolated processes become more effective and efficient. Examples include word processing tools, enterprise resource planning (ERP) systems, manufacturing execution systems (MES), and artificial intelligence solutions like translation tools and large language models.

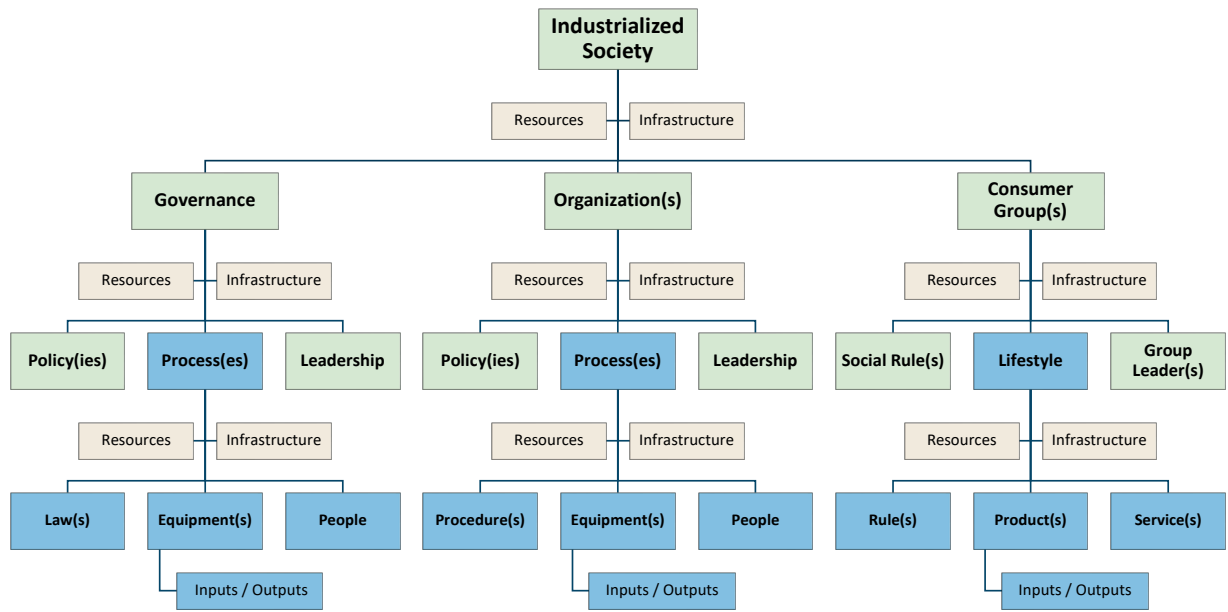

*Figure 3 **Digitalization:** Transformation at process level, including controls and human interface.*

Digitalization enhances the value delivered by digital devices to a new level. A digital camera just captures and stores a picture. Digitalized photography allows enhancing, printing, sharing, making albums, tagging, searching, and doing a host of things that were not possible earlier. It opens new possibilities. Digitalization is having a serious impact on work and personal lives, with the internet, entertainment, gaming, cell phones, self-driving vacuum cleaners, etc.

For a company or an individual, a process digitalization is essentially a project comprising of multiple tasks, that together make that process more efficient, effective, and robust, using digital devices. Unfortunately, many company managers believe that to be the end state of digital transformation – process simplification automation for productivity and cost, which is just the beginning of a serious change.

In cars, digitalization is more obvious to a consumer than digitization under the hood, with things like a digital dashboard to present all relevant information regarding the car's status, automation solutions for lights and windscreen wipers, remote access solutions, navigation systems, cruise control, and more. Some other, not as obvious, digitalization solutions address car maintenance challenges. Automation and digital controls are obvious examples of digitalization for car manufacturers.

## 3.3   Digital Transformation

Digitalization of standalone processes, as discussed above, uses new workflows and ways to control inputs and outputs. This offers an opportunity to connect all digitalized processes in the company leading to the transformation of the value stream and in turn the entire operation (middle vertical in Fig. 4). This is the authors view on Digital Transformation.

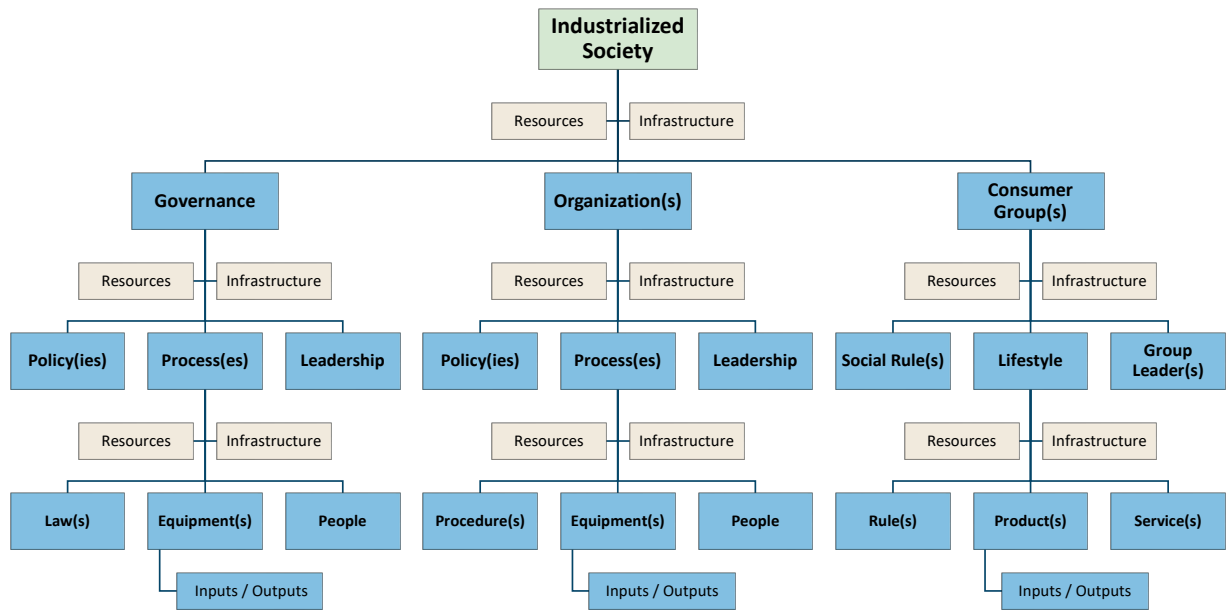

*Figure 4* **Digital Transformation**: *Transformation at organization, governance, or consumer group level*

At this stage, the leadership needs to become the digitally active change agent in business transformation, ensuring digital enforcement of company policies, and eventually achieving a state of 'Digital Transformation' at the value creation level. When technology and business leaders come together, they can see this clearly and actively drive appropriate changes.

The digital transformation on the consumer side (see Fig. 4 right) is driven by the big IT players and has already developed a lot further in comparison to the digital transformation of organizations. Examples include social media platforms, various cloud-based tools, online gaming, music and video streaming platforms, virtual assistants, and devices for smart homes, as well as the current-day smart city activities. Most of those technologies by themselves simplify isolated processes and should therefore be seen as digitalization solutions. But with establishing connections between those solutions the consumer market leans towards digital transformation, for example, multi-modality mobility solutions or seamless vacation planning.

Although automobiles are getting more and more digitalized, to a point of allowing self-driving cars, the digital transformation of mobility is not obvious up to the moment. It will come in different shapes and forms for different entities in the ecosystem:

- From a technology standpoint, machine-readable vehicle-to-vehicle and vehicle-to-infrastructure communication could become the basis for digitally transformed self-driving cars.
- From a car manufacturer standpoint, it would mean the integration of generative designs, resource planning, manufacturing automation, human resource management, marketing and sales, and new business models. This could extend into automatic supply chain

  management with standardized communication to enable collaborative design, manufacturing planning, pricing, custom order tracking, cash flow, etc.
- From a user standpoint (both B2B and B2C) the integration of the car into multi-modality mobility solutions must become reality. A traveler should be able to go from point A to point B with a single click or voice command, rather than multiple disjointed websites requiring separate accounts/payments and manually ensuring connections and schedule margins.

Digital transformation will bring all individual and public transportation solutions seamlessly together to optimize travel from an efficiency, booking (so-called deep integration), comfort, duration, and sustainability viewpoint – not only regarding people transportation but also luggage or goods transportation or even farm produce. It also allows enhanced mobility and modality research allowing, for example, to enable enhanced parking lot search and city planning.

Considering the necessary scope of change to achieve a level of digital transformation as pictured above, it requires sustained leadership commitment over years with continuous investment into a prioritized roadmap for technology and talent development. It also involves changes in leadership style and mindset. No wonder, it is referred to as a journey, to be led by organization leaders.

For most companies this digital transformation has just started with rudimentary applications of the Internet of Things (IoT) and digital twins, leveraging machine-to-machine communication systems like OPC UA for productivity and cost benefits [15-17].

In summary, for a company, a consumer group, an individual, or governance, digital transformation should be seen as a journey, wherein processes are first digitalized and connected to make the entire operation more effective and efficient. This becomes simpler and more affordable if the connectivity was planned into the digitalization from the start.

## 3.4 Digital Transfiguration

Fig. 5 shows that the transformation does not end with a fully digital organization. There is another level at the ecosystem, which requires a digital connection between state-level governance, companies and industrial sectors, and diverse consumer groups. This is the stage where the industrialized society will gain a new configuration, a new persona that today can only be imagined in parts. The seamless interconnection of mobility, homes, healthcare, education, services, and products will lead to a profound lifestyle change. It will be more than digital transformation as generally discussed today in leadership conversations and published media.

The authors suggest calling it **"Digital Transfiguration"** as it re-configures social norms and behaviors.

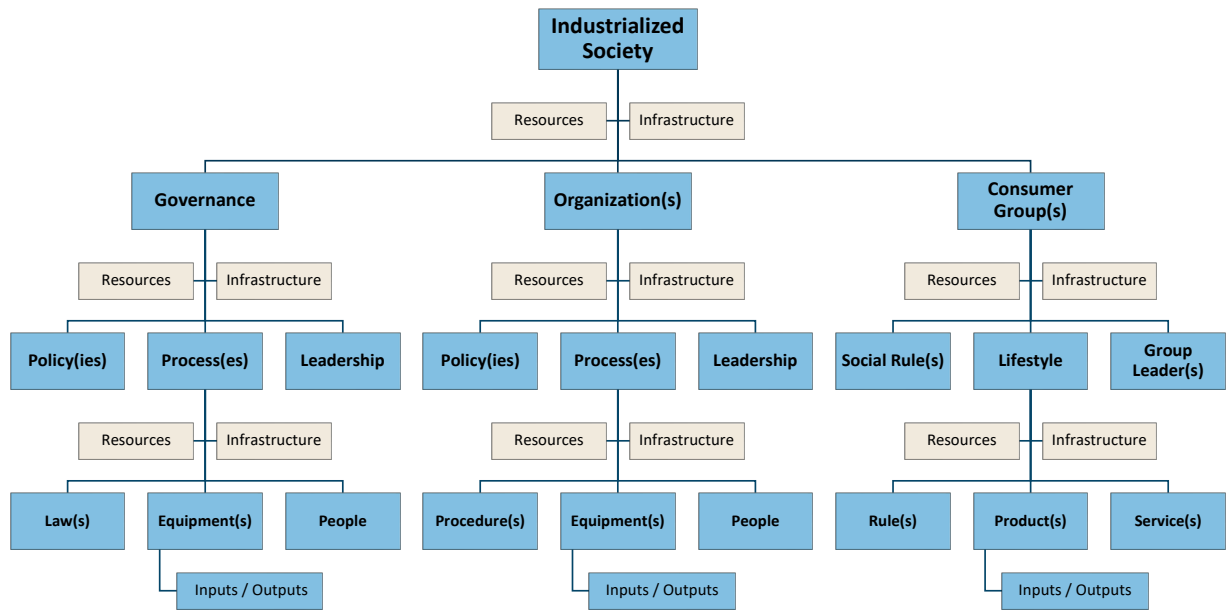

*Figure 5* **Digital Transfiguration:** *Transformation at society level*

To better understand this concept the discussion of the future of mobility will be continued. Digital transformation will lead to the seamless integration of all modalities into self-driving cars. But with digital transfiguration, even more, interesting concepts might become reality. Say a person wants to go to a country requiring a visa. During the digitally transfigured booking process, the system automatically makes the person aware and one mouse click later all the relevant information is submitted, paid and the visa application processed. Which then gets added to the person's digital global entry passport, which already works with facial recognition. To get to the airport the person just steps out of the house and gets into the self-driving car waiting. The luggage is checked through from the car to the airplane and all the way to the final destination using some mobility automation. This all leads to a drastic reduction in the number of cars, parking real-estate, and carbon footprint required for sustainable mobility. Digital transfiguration is the basis for truly revolutionizing the mobility lifestyle, in some sense achieving the vision of Society 5.0 given by Japan [21]. This profound change to the social fabric is more than a journey.

It is more like a **"voyage"**, that brings meaning to all the digital transformation of various industrial sectors and consumer groups. One that truly impacts human lifestyle in a favorable way. And when done responsibly, can ensure sustainability goals as well.

Here is a scenario of a socially beneficial transfigured system. An automobile accident triggers an emergency response, where the ambulance comes prepared with equipment required to stabilize the patient, whose condition was simultaneously transmitted by a wearable device. The ambulance takes the patient directly to the most appropriate location for treatment, with complete access to the patient's medical condition prior to the accident. The correct physician is engaged via augmented reality right from the moment of triage. Any surgical implants are manufactured using 3D printers in the hospital, designed to fit the individual human body, whose complete digital model already existed in the records. Medicine gets delivered through drones to wherever the patient is. The insurance firm gets activated with a data feed from the car and the doctor. Family and friends get notified automatically based on custom protocols programmed in individual devices. The non-availability of the individual triggers appropriate substitute protocol at their workplace, informing all parties impacted based on calendar schedules. And so on. Opportunities for the smooth handling of such a contingency are endless here.

All this is assuming that people still need to move. Perhaps virtual transportation might make automobiles or personal drones a thing for history books.

## 3.5 Taxonomy

Based on this model and understanding, Table 1 presents a consistent set of definitions through the four building blocks of digital penetration, that ties actions with outcomes.

*Table 1 **Digital Penetration Taxonomy:** a consistent set of definitions through the four building blocks that connects actions with results.*

| | |
|---|---|
| **Digital** | Something using discrete digits.<br>Often string of binary numbers (0s and 1s)<br>to represent an analog signal, message, picture, instructions, and program. |
| **Digitization** | Changes to value-creating devices or equipment,<br>enabled by **digital** storage,<br>for the universality of application. |
| **Digitalization** | Changes to an isolated process,<br>enabled by **digitization** of devices, equipment, and human-machine interface,<br>for simplification or automation. |
| **Digital Transformation** | Changes in value creation or consumption,<br>through the integration of **digitalized** processes and **digitized** devices,<br>for enhanced business value and consumer experience. |
| **Digital Transfiguration** | Changes in industrialized society,<br>from massive integration of all **digitally transformed** entities in the ecosystem,<br>to enable the human pursuit of self-actualization. |

## 3.6 Characteristics of Digital Penetration

Digital penetration started with the adoption of digital equipment (Digitization), to digital processes (Digitalization), to processes integration at the value creation level (Digital Transformation), all the way to having the entire creation-consumption value cycle as digital, creating meaningful consumer experience (Digital Transfiguration).

Digitization is the core of the third industrial revolution. Enhanced digitalization and digital transformation are key to the fourth industrial revolution [22], and digital transfiguration is what provides meaning to the fourth revolution. This is depicted in Fig. 6.

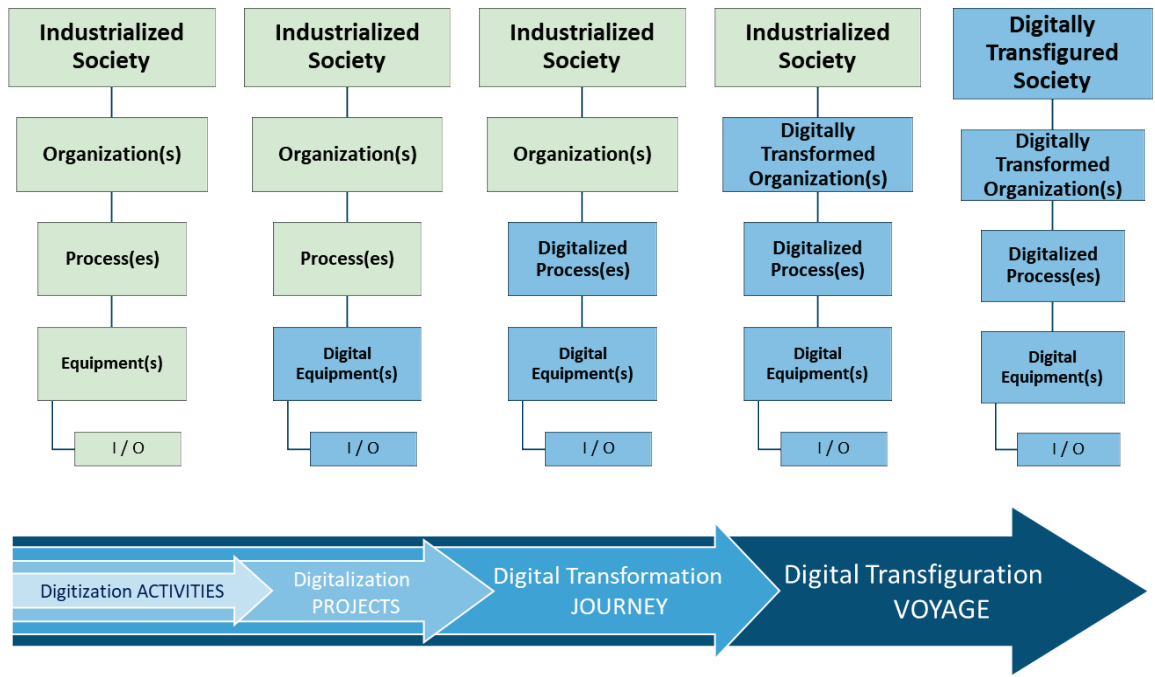

*Figure 6 **Digital Penetration** - From the industrialized society, established during the 2nd industrial revolution, to the digitally transfigured society.*

In some sense, digitization can be seen as the digital transformation of a device or equipment. And equally correctly digital transfiguration can be seen as the digitization of the entire ecosystem. No matter what, the interesting thing is digital is the transforming catalyst in the process that changes something else. Digital is only a means to meaningful living, and the focus must be on the pursuit of purpose (social) more than the means (digital).

Another important aspect to appreciate is that these are not four discrete levels. The change happens gradually, albeit rapidly. Just like different companies are at different stages along the voyage, different departments within a company can be at different stages of their journey.

In the following the various facets of digital penetration will be explored addressing the need for enabling infrastructure, required skills, and additional governance.

### 3.6.1 Evolution of Enabling Infrastructure

The most significant factor for a successful transformation is the infrastructure. And it scales up exponentially with digital penetration, requiring new abstraction layers, shifting the burden stepwise from changes in hardware to software and communication.

**Digitization** requires no or little external infrastructure, as it is mainly a hardware effort, built-in as a functional part of the device (microelectronics or storage media). But it does require the development of rules for how digitized data is encoded and stored in binary form. In addition, devices need a source of energy. Equipment manufacturers are responsible for this.

**Digitalization** requires new infrastructure levels on top of the data encoding binary infrastructure, like standardized computer operating systems, software development platforms, as well as connectivity along the process, such as servers, wired and wireless networks, the Internet, and the cloud. One of the requirements at this point is syntactic interoperability. Information and communication technology (ICT) companies provide this, managed by the IT departments of the host companies.

**Digital Transformation** requires connectivity to go across value streams and supply chains, which may be global. The hardware infrastructure required stays about the same as for digitalization, it just needs to be adapted to the increased communication and data processing needs. In contrast, a completely new architecture level is needed on the software side. At this level of connectivity, semantic interoperability becomes a prerequisite – data must become machine-readable information that can be understood automatically by any computer. Just like a human can understand every text written. One of the challenges today is that every device, system, or software package comes with its proprietary data formats and interfaces. Almost every software needs to be programmed explicitly to understand the data of the other systems. To enable serious digital transformation along the supply chain, an infrastructure is needed that allows data transparency, machine readability, and data sovereignty. Company CIOs need to understand this and lead this initiative to make it successful.

**Digital Transfiguration** requires connectivity to go across all industrial sectors, government entities, and geographical regions of the global consumer society. This requires conceiving a massive new software infrastructure that allows appropriate data transparency, machine readability, and data sovereignty. The semantic interoperability needs to be based on a global ontology with international public infrastructure. Universal machine readability requires a radically new implementation of device/computer operating system clarifying data ownership and sovereignty while ensuring data safety and security and broad acceptance through novel business models. While digitization, digitalization, and digital transformation were somewhat possible using proprietary solutions – digital transfiguration will not. The idea of digital transfiguration falls within the purview of the United Nations as described in Chapter 1, Article 1, Purpose 3 ("To achieve international cooperation in solving international problems of an economic, social, cultural, or humanitarian character, …") and 4 ("To be a center for harmonizing the actions of nations in the attainment of these common ends.") [23].

Clearly, the gaps in infrastructure are slowing the journeys of the companies and the voyage of society and holding humanity back from realizing the full potential of even mature digital technologies. Just like the slowest part of international travel is connections at passport control and airport security, despite the superfast trains and airplanes.

### 3.6.2   Required Digital Skills

Another important requirement for massive change is the identification of new skills and progressive talent development. The need for other resources such as funding and facilities should not be downplayed, but skills are the most significant resource. Considering how far humanity has come over the last few decades, the technical skills for early-stage penetration are well-developed, whereas the final stages are not yet clear.

**Digitization** requires skills like physics, electrical engineering, and basics in data encoding. Those competencies are understood for decades.

**Digitalization** requires skills like computer engineering, IT security, and data engineering. In addition, it requires a good understanding of workflow, lean process, value stream mapping, IT networking, feedback loops and controls, and automation. These have also developed over the last few decades, mostly in emerging economies, with enough capacity to meet the global needs of tomorrow.

Consumer acceptance becomes important at this stage and a certain level of digital competency is necessary for all of us. The third revolution had created a major rift between digitally literate Gen-X and baby boomers. With the advent of natural language user interfaces and touch-friendly devices with graphics, this is getting easier. Yet care must be taken to carry the entire society along by reducing the learning burden on the consumers, particularly the aging population.

**Digital Transformation** requires skills like cloud security, data sciences, information science and engineering, information-to-value engineering, etc. The challenge is the combination of subject matter and digital skills within an individual. Most companies must choose between training an IT guy on OT or an OT person on IT or making a multi-disciplinary team that collaborates closely, across domain knowledge specialists (the various engineering disciplines, business, marketing, sales, purchasing, controlling, human resources, training, …). The softer side of skill set gets very close to major organizational change and perhaps most of the well-established tricks would work when leaders begin to see new possibilities. At this stage unlearning some of the Industry 2.0 and 3.0 business practices may be more important, which allows an open mind to embrace the new reality.

**Digital Transfiguration** skills requirements will emerge in due course. Early indications are that it requires skills like information ethics, data rights, and sovereignty, empathy and compassion, social-psychological aspects of collaboration across society, appreciation of human drivers across geographic areas, knowledge domains, and work cultures. Knowledge owners in government, enterprises, and consumer groups must learn to work collectively – on a global scale, addressing human issues. An additional key element will be keeping it under control in the public's general interest. Governance and public policy skills get important here, with the acceptance that there will be new possibilities from which there was no precedence or experience to learn from. This final step from Digital Transformation to Transfiguration will be possible only when international bodies can appreciate the possible scenarios and drive them.

Invariant skills required throughout, include empathy, cognitive skills, and creative problem-solving. One must practice the art of asking questions that start with why not, what if, and how about. Boldly exploring without boundaries, including diverse viewpoints, accepting setbacks, and discussing ethical and moral dilemmas, are all important, just because the change is fast. The creative part becomes more interdisciplinary and visionary as humans progress toward transfiguration.

Within the unified model, the skills ought to be stackable, just like in any organization.

### 3.6.3 Additional Governance and Controls

Checks and balances are an important aspect for any change and progress. Despite all regulations, serious side effects happen from massive changes, primarily from the scientific understanding that evolves with experience, leading to revisions in the control process. Emissions, noise, asbestos, vaping, etc. are examples and digital is no exception. Data is viewed as material and poses similar challenges as it flows across boundaries.

**Digitization** requires no external governance or controls. These are built into the device as a part of product specification and functional limitations.

**Digitalization** requires process control and cyber security using digital logic. Quality control must include software debugging and mistake proofing. It also calls for additional ethical guidance [24] to meet the challenges of digital outcomes. Any built-in AI must follow ethical guidelines and principles.

**Digital Transformation** requires semantic interoperability and governing principles for digital operations. This is the stage where company leaders might need to develop policy guidelines for digital progress. At this stage, consideration should be given to sustainability, sustainable development, and circular economy. A company must consider having a Digital Transformation Review Board comprising of responsible individuals with experiential and cognitive diversity.

**Digital Transfiguration** requires global semantic interoperability and data sovereignty. This is where governments and industry leaders need to develop federal policies and regulations under the umbrella of the United Nations. Currently, digital transfiguration is happening serendipitously without centralized leadership. This gap is serious and must be closed.

Within the unified model, the digitally implemented procedures, digitalized compliance with company policies, and digitalized governance of the ecosystem must be aligned from the top down. Since each of these will be evolving over the next decade, annual reviews of this alignment become an important strategic leadership action.

### 3.6.4 Overview

Taking a deeper look at the stages of digital penetration it becomes clear that technological innovations are at the bottom of it or just at the beginning and the resulting revolution of our society is at the top. This sustained change through various stages is not to be confused with rapid continuous improvement within a stage. It needs to be viewed as creating new value propositions not previously possible, in addition to making existing systems better. Table 2 summarizes these stages to provide an overview.

*Table 2 Overview of the stages of digital penetration*

| | | | | |
|---|---|---|---|---|
| **What is it?** | | | Digital Transfiguration Voyage | |
| | | Digital Transformation Journey | | |
| | Digitalization Project | | | |
| | Digitization Task | | | |
| **What Transforms?** | Devices & Equipment, Data Storage, Input and outputs | Processes & Workflows, Span and Mode of Controls | Business and Consumer Behavior | Entire Society |
| **Who owns** | Team Lead | Project Lead | Company Leadership | Serendipitous (Current Gap) |
| **What Improves?** | Performance & Capabilities | Processes Efficiency and Effectiveness | Factory, Home, Vehicle, Hospital, Coursework, … | Supply chain, City, Mobility, Healthcare, Education, … and their interactions |
| **What Happens?** | Universal Storage | Simplification & Automation | Enhanced Value Creation | Enhanced Lifestyle, and Society |
| **Infrastructure** | Built into the Device | Cloud, Syntactic Interoperability, OS and Wi-Fi, AI/ML | Semantic Interoperability | Global Ontology |
| **Key Skills** | Fundamental Disciplines, Coding, Logic, databases | Computer Engineering, IT Security, Lean Principles | OT and IT Skills Combination, Ethics Organization Behavior | Ethics, Sovereignty, Public Policy |
| | Empathy, Cognition, Creative Problem Solving, Curiosity, Listening, and Collaborating | | | |
| **Governance** | Built into the Device | Process Control, Cyber Security | Principles for Digital Operations | Federal Policies and Regulations |
| **Car Example** | Under the Car Hood | Driver-car Interface | Multi-mode Mobility | Frictionless Movement |

This table provides a distinction between digitization, digitalization, digital transformation, and digital transfiguration, as well as the need for connectivity infrastructure, additional skills, and governance at various stages of change management. It provides an outline for a transformation roadmap helping understand the current state, see the future state, and facilitate technology, infrastructural, policy, as well as skill gaps along the way. The holistic view of digital penetration brings out the distinction between the digital transformation of a company from the digital transfiguration of the ecosystem, which connects the company, the consumers, and regulatory bodies. The table clearly shows that digital transformation is not a definable destination, and it is also not about digital technology. It is about humans, actions, behavior, and values. Technology is merely a catalyst that must be carefully adopted.

## 3.7 Leading Meaningful Change

Having discussed a unified model of transformation, it should be possible to see the status of the voyage to digital transfiguration. Irrespective of how people perceive themselves, either as value-creators or value-consumers, they will always have a very limited view of changes happening around them. The internal capacity to understand and comprehend all of it is not evolving as fast as the external changes witnessed. For most people, the confusion is thus natural in such a situation. The authors hope the model above clarifies the landscape to some extent and helps guide the transformation in a logically defined manner.

- Digital technology developers can see themselves creating effective digital products and interfaces and help identify the need for digital infrastructure.
- Managers can see the gaps in the digitalization of a process and focus on selecting appropriate digital technology.
- Organizations can systematically map and integrate various processes.
- Service groups can better understand the consumers and connectivity to address their needs.
- Regulatory bodies can map out the ecosystem and identify necessary infrastructural needs and controls required for the safe and effective insertion of digital means between creators and consumers.
- Workforce development bodies, corporate talent managers, and universities can build better curricula for various learner segments.

For leaders chartered with transformation, here are a few things that can help reduce risk and anxiety through the change process.

### 3.7.1 Think Purpose, Not Project

Digital Transfiguration requires mental preparation in the context of a long voyage, that requires sustained change effort. This needs to be used to guide a strategic plan and then a tactical project. Some may call it a journey if they prefer to plan for just the next few steps. At any point, everybody should be able to assess what parts of the industrialized society are in their control, and what must be adapted to. Focus needs to be on adopting digital processes and devices that are in control in line with adapting to the digital transformation around everybody in the ecosystem as shown in Fig. 6. Slowly and steadily the wave of change upwards needs to be pushed.

In the present circumstance, generally, the managers understand their process and can effectively delegate digitalization tasks to digitalize processes. However, for them to claim that they are on a digital transformation journey they must see their impact on the creation side of the ecosystem business. And to claim that they are on a voyage, they must be able to see its impact on consumer society. That thought process has a profound effect on successful change.

### 3.7.2   Accept the Uncertainty of the End State

There is very little value in demanding a precise understanding of the end state because there is none. The good and bad news is that digital technology gets obsolete fast. So, the business case for digitization and digitalization needs to be short to mid-term, with digital transformation as a long-term objective, and digital transfiguration as the purpose. Along the voyage an open mind and adjustments are key. Faith in the team is more important than confidence in ROI calculations. It will be hard to have a good estimate of returns from a project, let alone the entire transformation.

The way to look at it is the cost of not doing is likely to be more than the cost of doing it.

### 3.7.3   Upskill the Team

Major transformation needs serious education of people at all levels. In terms of why and what –the intent must be clear and plans for the upcoming changes must be shared openly. Employees will embrace it better when they are a part of it. And they ought to be, or there will be no transformation. In terms of how – digital skills are significantly different and important for success. So, investing in learning and development initiatives targeted toward a vision of the new state becomes critical.

### 3.7.4   Think Ecosystem

Go beyond proprietary solutions for short-term gains. History shows that proprietary digitization solutions eventually get replaced by open solutions building the necessary infrastructure so that everybody can benefit. Digital transformation is not a product that anyone can buy with sufficient financial resources and derive a competitive advantage. It requires serious collaboration to create value for everyone. Accepting the data exchange standards and using them is the key. There is a lot more value in working together under the new rules, many of them have yet to be written.

## 4   To Be Continued

Authors used the design thinking approach to model the industrialized ecosystem and understand the changes ensuing from digital penetration. Since this is a rapidly evolving scenario, they will refrain from writing 'conclusions' or 'closing remarks' in this manuscript. They can only think of a 'continuation'.

Just like in the early 19th century, different people had different views of what horseless carriages (cars) would be like, today humans have different views of what digitally transformed mobility or transfigured society will be like. Driverless cars and multi-modality mobility will become reality. This is coming along with Digital Transformation. It is concurrently asking for 5G+ infrastructure, vehicle-to-vehicle communication, standards for data exchange enabling machine readability, self-driving ethics, new car-sharing business models, renewed roads, AI-based driving regulations, novel insurance, and liability policies, and much more. A human driver might become the biggest risk in people's movement. How mobility will look in 50 years is difficult or even impossible to predict. But it will be radically different, with entirely new infrastructure, businesses, and governance. The following scenario can be imagined: no more roads if mobility moves on to personal autonomous drones, off-ground hoverboards, self-regulating rail-road systems, or a life-like metaverse. Almost a courageous Bertha Benz of the 2030s can be seen lifting off in a personal drone across the Sahara Desert, to prove a point. Similar is the case for changes in every other sector, and the eventual collapse of domains into the lifestyle. Still, humans must try to understand and lead it with a purpose, while leveraging lessons from prior transformations.

The unified model proposed here captures how value is created and consumed using digital devices and digital connectivity. It is humans, actions, behavior, and values that are transforming along the way. The model forms the basis for planning a meaningful digital voyage and continuation of the human quest to climb Maslow's hierarchy of needs. The reach of digital transfiguration becomes clear looking into the charter of the United Nations [23] as it falls within the purview as described in Chapter 1, Article 1, Purpose 3 ("To achieve international cooperation in solving international problems of an economic, social, cultural, or humanitarian character, …") and Purpose 4 ("To be a center for harmonizing the actions of nations in the attainment of these common ends.").

There is another confluence on the horizon. The penetration of biological materials into the physical systems will have the power to redefine human engagement with devices and ecological systems. All the revolutions up until now addressed the economic ecosystem and drove the accumulation of personal wealth and prosperity. They came with several side effects to the ecological ecosystem that are now becoming clear in terms of material waste, climate change, and the extinction of certain species. The digital-physical-biological confluence shows promise to transfigure responsibly and sustainably. That could be the fifth revolution. In such a situation, this model will further evolve with experience along the way and authors work hard to remain curious, always listening with an open mind, ready to engage.

# Supplementary Material

## Prior Art Review

George Westerman et al. defined digital transformation in 2011 as the use of technology to radically improve the performance or reach of enterprises [1]. This report covers a global study of how 157 executives in 50 large traditional companies that are managing and benefiting from digital transformation. The central theme was that successful digital transformation comes not from implementing new technologies but from transforming an organization to take advantage of the possibilities that new technologies provide. Major digital transformation initiatives should be centered on re-envisioning customer experience, operational processes, and business models. The study was focused on the 'how' more than the 'what.' Their position can be understood since the thinking tends to be linear from prior revolutions. But the 'what' becomes important when the social acceptance side of the equation is considered.

Lars Hamberg studied two concepts, digital transformation and digital maturity, in this recent 2022 study [6]. He considered three inputs: a structured review of scholarly literature where both concepts occur, an AI-based study of both concepts and a qualitative study with input from 177 experts on digital transformation. His findings confirm the conceptual ambiguity of both concepts, digital transformation and maturity. The paper also highlights possible misconceptions in literature and suggests that the successive changes between states of digital maturity are non-linear, as opposed to linear, and multi-directional, as opposed to unidirectional. The authors believe the missing piece is the consumption side.

Several studies and models that appeared in literature over the last 20 years, have been thoroughly reviewed by a few scholars. The most recent one comes from Maria Feliciano-Cestro et.al. [7] covering 272 articles published between 2002 and 2022. Most studies in this area were found to be empirical in nature, exploratory research, or conceptual articles. Based on an exhaustive review, they indicated the direction of future research: focus on empirically validating emergent theoretical constructs regarding this phenomenon or providing empirical evidence concerning the relationships between the various human and non-human components of digital transformation. This study certainly reinforced the authors' belief to bring humans into the equation as well as open the mind to possible items other than business and consumer.

Vial reviewed 282 papers to understand specific aspects of digital transformation and felt that there is still a lack of a comprehensive portrait of its nature and implications [5]. He found 28 sources offering 23 unique definitions. He used systematic decomposition to create a new definition of digital transformation – a process that aims to improve an entity by triggering significant changes to its properties through combinations of information, computing, communication, and connectivity technologies.

Of all the studies, it was clear that Vial had identified the gap or need to include society in the digital transformation model. His framework of digital transformation is built upon relationships that emerged through his analysis across eight overarching building blocks describing DX as a process where technologies play a central role in the creation as well as the reinforcement of disruptions taking place at the society and industry levels. These disruptions trigger strategic responses from the part of organizations, which occupy a central place in DX literature. Organizations use digital technologies to alter the value creation paths they have previously relied upon to remain competitive. To that end, they must implement structural changes and overcome barriers that hinder their transformation effort. These changes lead to positive impacts for organizations as well as, in some instances, for individuals and society, although they can also be associated with undesirable outcomes. This highly referenced work validated the direction of this study.

Another well-cited paper was authored by Verhoef in 2019 [8]. According to their study, digital transformation occurs in response to changes in digital technologies, increasing digital competition, and resulting in digital customer behavior. Despite identifying consumer behavior as an important aspect, they did not go past the three well-established stages for digital transformation: digitization, digitalization, and digital transformation. They identified 21 open research questions under 5 subtopics: digital transformation, digital resources, organization structure, digital growth strategies, and metrics and goals.

Zhang et al. mathematically modeled survey data from 180 Chinese companies to study the impact of IT infrastructure on digital transformation [9]. The model provides theoretical guidance on deriving digital transformation performance from IT infrastructure investments. This topic is the linchpin of transformation as seen in so many industries – automotive, communication, space, etc.

There is also very little mention of governance/controls/policies etc. in such a vast body of knowledge. Most of the literature studied focused on business entity management - strategic, innovation, operations, marketing, and information systems. There are very few studies alluding to policymaking as detailed in the following.

Arewa touched on the topic of regulation, looking at both the positive and negative sides of disruption [10]. Digital transformation may, along with other factors, intensify existing societal divides, lead to greater inequality in many places, and contribute to a scarcity of opportunities for many people. Digital economy policies must take account of the requirements of an economy permeated with the potential adverse effects of digital transformation. The paper identifies that Amazon has been at the center of digital transformation for almost three decades and exemplifies the benefits of digital transformation as well as sources of digital economy discontent. The scope of potential disruption may extend far beyond the market areas within which such companies operate; including how humans interact, work, play, live, and regulate. Digital economy transformation requires innovative approaches to regulation in a complex arena of multiple and potentially overlapping areas. Later in this paper, it will be shown that the digital transformation of companies must be differentiated by the digital transformation of user group(s) and it will become clear that Amazon's philosophy of reducing friction led to consumer acceptance of novel approaches that got them into the households – the DX of the user groups. Their model has gone beyond DX into social sectors but has never been formally mapped.

Mukesh observed that while the scholarly attention of DX is mainly towards entrepreneurial innovation in pure digital businesses at the organizational and individual levels, it has implications at a higher level of aggregation, like the regional and national levels [11]. His concept of digital transformation includes digital business model innovation and entrepreneurial innovation at the country level. Using fuzzy-set qualitative comparative analysis for a sample of 55 countries, he explored the causal configurations explaining the implication of digital transformation. The result brings out the need for digital infrastructure, digital regulatory framework, digital skills, and risk tolerance. The conceptual framework presented, however, does not provide any connectivity or nesting of these requirements.

These gaps are quite apparent in other studies as well and helped the authors to define the scope of the study discussed in this paper: Develop a single holistic model that brings together all these: value creation, value consumption, infrastructure, skills, governance, drivers, enablers, constraints, and limitations.